\documentclass[aps]{article}
\usepackage{geometry}                
\usepackage{graphicx}

\usepackage{amssymb}
\usepackage{epstopdf}
\usepackage{amsmath}

\usepackage{amsfonts}
\usepackage{bm}
\usepackage{hyperref}

\DeclareMathAlphabet{\mathpzc}{OT1}{pzc}{m}{it}

\title{Thermalization in Quantum Systems: An Emergent Approach}
\author{Clifford Chafin\\\ \small{Department of Physics, North Carolina State University, Raleigh, NC 27695} \thanks{cechafin@ncsu.edu}}

\begin{document}
\maketitle
\begin{abstract}
The problems with an emergent approach to quantum statistical mechanics are discussed and shown to  follow from some of the same sources as those of quantum measurement.  A wavefunction of an N atom solid is described in the ground and excited eigenstates with explicit modifications for phonons.  
Using the particular subclass of wavefunctions that can correspond to classical solids we investigate the localization properties of atomic centers of mass motion and contrast it with more general linear combinations of phonon states.  
The effectively large mass of longer modes means that localization present in the ground state persists on excitation of the material by macroscopic coherent disturbances.  The ``thermalization'' that arises then follows from the long term well defined motion of these localized peaks in their 3N dimensional harmonic wells in the same fashion as that of a classical solid in phase space.  Thermal production of photons then create an internal radiation field and provides the first dynamical derivation of the Planck distribution from material motions.  Significantly, this approach resolves a long standing paradox of thermalization of many body quantum systems from Schr\"{o}dinger dynamics alone.  
\end{abstract}

Thermalization in classical mechanics has a long history that had its important beginnings in the kinetic theory of gases.  So unready was the scientific community for these ideas that they appeared multiple times and were lost before reaching even the level of acceptance for publication and large scale debate.  These arguments were long and bitter with both sides 
stellated with celebrated scientists.  The very introduction of statistical methods into the subject by Maxwell was resisted and mocked by mathematicians.  Even with Boltzmann's improvements, they battled and challenged by Poincare, Zermelo and other luminaries.  By 1906, Boltzmann was so depressed, in part by the lack of acceptance of his work, that he had written a condensed version of his work so it would not be completely lost \cite{Boltzmann, Jeans} and committed suicide.  Ironically, three years later his work had been validated and was completely accepted by even his strongest detractors.  Notions such as entropy and the second law became practically fundamental.  The Gibbs' paradox remains as an unresolvable problem in the classical theory but classical statistical mechanics is still very useful in certain applications.  In the last twenty years, the classical subject has seen a huge resurgence in topics involving fluctuations, reversibility \cite{Jarz, Cohen} and rigorous bounds on equilibration \cite{Villani}.  

The quantum version of thermodynamics has never been as conceptually clean.  The Gibbs' paradox vanishes by the quantum symmetries of bosons and fermions but bigger problems persist.  Unlike the classical case, such systems seem to have no universal tendency to thermalize even on long time scales.  Classical systems wander around on a submanifold in phase space defined by the conserved quantities.  Ergodicity is not a viable solution to the problem of classical averaging because of the long time scales involved in sampling the manifold but, nevertheless, there is a very large subset of most probable states that manifest such a statistical representation of interesting quantities.  Quantum systems are governed by wavefunctions and, once the distribution of eigenstates are given, this never changes.  Classically, fluctuations arise from the random clustering of particles in phase space but, in the quantum case, energy fluctuations depend directly on the spread of energy eigenstates used in the formation of the state.  Most distressingly, the ``ensembles'' we use in the microcanonical formulations of quantum states are energy eigenstates have the same energy so any wavefunction built solely of these has zero energy fluctuation.  

Of course, we do calculate fluctuations in quantum statistical mechanics.  This is done by looking at subsystems thermostated by reservoirs.  One can then calculate the probability of an energy fluctuation by the subsystem in the distribution.  Some big problems are hidden by this approach.  First there is no natural way to ``partition'' a general many body wavefunction into such ``parcels.''  Indeed, the class of many body wavefunctions that correspond to classical objects is rather special and far from what we would consider to be eigenstates \cite{Chafin-I}.  Next, the nature of calling such a variation a ``fluctuation'' should be suspect.  Quantum discussions often have a rather formal character and so many conceptual inadequacies exist in the subject that a low bar for their acceptance and interpretation has become part of the culture as a necessity to the ``shut up and calculate'' culture promoted by Feynman and others.  A classical fluctuation exists as a variation not only in space but in time.  We can track a typical state in classical phase space and observe the size and rate of them without the need for any external stochastic forces.  Such a condition is missing from quantum considerations.  There is no ``typical'' thermal wavefunction for us to similarly evolve.  Ensembles are formally treated so it is not surprising we have no idea how to extend the ideas of the recent classical statistical mechanical revolution to the quantum cases (beyond more formal manipulations of already formal constructions).  

Some early work on the problem was very much built on the assumptions that a dynamic resolution was possible.  Einstein and Hopf \cite{EinsteinH} investigated the conditions on a quantum oscillator moving in a radiative field and found that if the distribution was Planckian then there should be no net force on it.  Einstein went on to say that a theory of radiation must show ``the impulses transmitted by the radiation field to matter lead to motions
that are in accordance with the theory of heat'' \cite{Einstein}.  Like so much of Einstein's conceptually inspired work on quantum mechanics and thermodynamics, they ultimately lost favor as more formulaic approaches that led quickly to experimentally verifiable results gained traction.  

It should be mentioned that there are occasional periods of recognition of this problem in physics.  Recently Popescu et al \cite{Popescu} have promoted the idea that coupling to the external world will drive the system to a thermal distribution automatically.  This is sort of parallel to the notions of decoherence whereby the external classical world drives an einselection process to produce quantum measurement statistics.  The eigenstate thermalization hypothesis and dynamical typicality seeks to show why the microcanonical ensemble gives accurate results given some constraints on the matrix elements of the observables \cite{Srednicki,Srednicki1, Bartsch}.  As in the case of the arguments against classical statistical mechanics in the 19th century, it is not surprising that these results are of a highly mathematical character.  The conceptual problems still remain yet an objection to them requires raising the mathematical bar again at a time when people have become even less swayed by consistency problems derived by thought experiments.  There is a sense of hopelessness in such methods and sometimes a heady optimism that formalism will solve our problems.  

Not surprisingly, this author has a negative view of both theories.  The quantum measurement statistics follow naturally from a worldview that states that there is a single many body wavefunction of the universe that is partitioned into a sparse set of classical domains.  Each of these has the kind of correlated shape, orientation, internal thermal states and location of condensed matter that is characteristic of the classical world.  While the reason for the genesis of such a condition is not entirely clear (but see \cite{Chafin-Qmeasurement} for a proposal using interacting small dust clusters), the persistence and independence of these are long lasting.  Quantum measurements then arise by further partitioning of them by interactions with delocalized lighter objects \cite{Chafin-II, Chafin-Qmeasurement}.  

Buoyed by such a realization, it seems reasonable to suggest that thermalization in quantum mechanics is best described by a typical wavefunction and its evolution.  Some barriers to this are that such ``classical states'' are far from true eigenstates.  Rotational symmetry is required of ground states so that a well defined orientation of a block requires it be a mixture of many higher angular momentum states.  The particular shape of a block requires some complicated mixture of hyperradial and hyperangular excitations.  Phonons arise as the eigenstates once we have accepted such a configuration as metastable but we now see these are clearly pseudo-eigenstates that have meaning only so long as these constraining conditions on the solid hold.  Even accepting that these are a useful persistent basis of quasiparticle excitations, one still has the problem of why arbitrary linear combinations of these are not allowed.  Distributions with large $\Delta E$ in the many body eigenstates give large fluctuation in time whereas one with $\Delta E=0$ have none.  We will show that the slow spreading property of larger mass bodies (and subregions of coherent thermal motion) comes to the rescue here, just as it does for the quantum measurement problem, and introduces a means of universal thermal limits and associated time varying fluctuations.  

We will see that there are three kinds of Poincare-like time scales to discuss.  The first for the time such a solid's vibrational state takes to recur.  The second is the time for the localization of atomic locations to fade and reestablish and, third, the time for the classicality of the system, i.e.\ the well defined shape, orientation, localization, velocity and rotation, to dissipate and reestablish itself.  The ``classicality'' of the system now establishes a well defined arrow of time since it gradually fades away and, not only will it not recur on any meaningful timescale, our very ability to maintain memory and function as somewhat reliable discrete state machines fades with it \cite{Chafin-Qmeasurement} and thus our consciousness to observe long time variations in thermodynamics is compromised by the same processes.  

\section{The Wavefunction of Solids: Phonons}

Let us pause to consider the distinction between classical and quantum systems and why classical systems must have fluctuations and why they have such universal character.  Classical particles are localized.  In a gas, these are ballistic and travel long distances relative to the nearest neighbor separation.  Cases that don't give typical fluctuations are very rare and unstable e.g.\ the case of a pair of countermoving particle lattices reflecting off each other in periodic motion.  Classical substates in quantum require localization as well.  When we look at the surface of a solid or liquid, it is never ``blurry.''  The phonon is often described as a quantum of sound but careful examination shows that the eigenstates are not shifting the well defined locations of atomic cores but delocalizing them in a stationary fashion.  The phonon also is calculated as a measure of displacement of the core from its equilibrium position in the lab frame not the more physically relevant location relative to its nearest neighbors.  

The ``ground state'' of our block can be considered a symmetrized product of atomic wavefunctions with their center of masses arranged in a periodic fashion.  This requires some elaboration.  This wavefunction in the core coordinates is a 3N dimensional object $\Psi\sim \sum_{\pi} \prod_{i=1}^{N} \psi(x_{i}-R_{\pi(i)})$ where $R_{i}$ are the 3D atom locations in the lattice, $\psi$ is a local peaked 3D wavefunction and $\pi(i)$ gives a permutation of the coordinate labels $\{i\}$ in the $x_{i}$.  Phonons are modifications of this that give a broadening of the core locations along the many body coordinate axes.  Multiple occupancies of phonons are higher modes with shorter wavelengths.  The 3N coordinate directions explain why there are 3N ``phonons'' in the system.  Such a state is implicitly symmetrized by the summation.  Note that this is a standing wave eigenstate so there is no current or time variation of the density.  Let us now make this more explicit.  

We can consider the motion of a single 3N dimensional peak corresponding to location $\tilde R=(R_{1},R_{2},\ldots R_{N})$ in the binding potential of the space defined by the two atom interactions and an artificially imposed external boundary.  This peak encodes all the locations of the classical lattice of atoms.  Quantum excitations form eigenstates that introduce nodes into these peaks of the wavefunction.  There are 3N normal mode directions corresponding to the classical normal modes along directions specified by the many body coordinates $X^{i}=\sum_{j}e_{j}^{(i)}x_{j}$ where the $e_{j}^{(i)}$ coefficients specify classical normal modes and the $x_{j}$ are the N-body coordinate labels.    Except for the center-of-mass (CM) coordinate of the system, the peak is bound in a harmonic potential of varying strengths along each such normal mode direction $X^{i}$.  The excitations give higher Hermite polynomials in their description with solutions $\psi_{n}(y)=H_{n}(y/a_{i})e^{-y^{2}/2a_{i}^{2}}$ where $a_{i}$ gives the scale of the excitations in 
the $y=X^{i}$ direction.  Locally at the many body coordinate location $\tilde R$ we have the solution $\Phi= \prod_{i} \psi_{n(i)}(X^{i}-R_{i})$ and $n$ gives the ``phonon occupancy'' of the mode in the $X^{i}$ direction.  To generate the full solution we have $\Psi=\sum_{\pi}  \Phi(X^{\pi(i)}-R_{\pi(i)})$ where the sum is over all permutations of the indices.  Such a state is considered to have ``n phonons'' in the ``$i$th mode''.  

\begin{figure}
  \begin{centering}
 \includegraphics[width=2in]{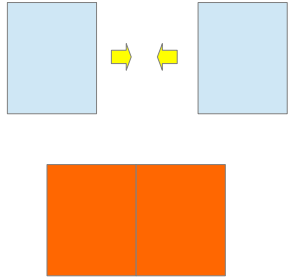}%
 \caption{\label{blocks} Two zero temperature blocks acquire thermal energy by a macroscopic collision.  The persistence of atomic localization over long times leads to a particular subset of phonon superpositions that can possess the universal properties of thermalization generally lacking for most quantum systems.}%
\end{centering}
 \end{figure}

Let us assume we have two blocks that are ``T=0'' states.  These are not ground states, even when at rest, due to the discussion above but we can expect them to maintain their localization of their center of mass and that of the atomic cores for very long times.  Assume they are now moving towards each other with velocity $v$ and then attach on contact as in fig.\ \ref{blocks}.  Naively we would expect the state to be a new larger block with some phonons that scatter around and equilibrate to some ensemble limit.  Certainly the new block is in some excited state and it is not an eigenstate.  By the reasoning above, the localization property of the cores is not going to change.  This means that we must have some combination of many body phonon eigenstates that has nonzero energy spread.  The initial impulse of acoustic energy from the collision will ricochet around and spread out.  In the classical case we expect equilibration to small scale ``thermal'' fluctuations for long times.  In the quantum case, we should ask how to best encode classical sound into our wavefunction.  Typically this would be done with coherent states.  Here let us both presume less and be less specific.  The oscillations will recoil off the boundaries and eventually cease to have any sense of classical coherence in that no 3D distribution of the acoustic energy will be tenable.  
Assuming no photon creation and destruction\footnote{Photons are fundamentally different than phonons here because they actually introduce new coordinate freedom to the system.  Phonons have a formal similarity with photons in their operator calculus but the dimensions of the descriptor never change.  This will be important below.}, we can describe the system with this single wavefunction $\Psi$.  The problems with this point of view will be visited shortly.  

\section{Equilibration}

The microcanonical ensemble is a formal statistical way to look at quantum thermal systems.  It states that, for a closed system, the ensemble average should be a sum over all states with the same energy (or in some tiny window about it).  This is problematic in several ways.  Firstly, the size of the energy window $E\pm dE$ is ambiguous.  If we make the window small too fast as N increases, the states become irregular and we get an non quasi-continuous distribution.  Secondly, we have only restricted energy not any of the other conserved quantities.  Third, any classical system lives exactly on a fixed energy sheet in phase space and, even though it cannot cover this hypersurface ergodically on any meaningful time scale, the averaged values are typically represented by most values traveling on it for most times.  A wavefunction can have the same energy and be a combination of eigenstates at different energies.  Indeed, this is the only way fluctuations can evolve.  Whatever this ensemble is encoding, it is not the evolution of a general many body wavefunction.  

Let us again consider the evolution of our fused blocks in fig.\ \ref{blocks}.  The long range energy density fluctuations will dissipate but the spread of energy eigenstates (in the new two block basis) is fixed.  This is the distribution that gives temporal fluctuations in the system.  If we are looking for a universal result as a function of the internal energy of the system then this picture is problematic.  For our solid block, the eigenstates are not single phonon modes but the full many body wavefunctions made of 3N phonon fields.  If we take a random phase mixture of all of these states of energy $E$ we still have an eigenstate of the same energy.  The success of many body quantum statistical mechanics suggests that such a state is dominant but with some spread of energy states around it to give time dependent fluctuations.  Furthermore, the localization intrinsic to classical condensed matter systems and the long lived quasi-bases corresponding to them seems to be crucial to the process of thermalization itself and to escape the apparent paradox of quantum systems remaining in a constant mixture of eigenstates.  

In the case of a classical gas (in greater than 1D), local equilibration to the Maxwell-Boltzmann distribution is very rapid.  The reflections due to the two body potential gives equilibration within a few collision times.  If we model a gas as made of localized, atom sized or smaller, wavepackets, as might arise from evaporation off a solid surface, such a dynamic also exists.  In a gas, the wavefunction starts to delocalize barring further interaction with the surface (and the yet unknown effect of photons) which introduces some complications but the eigenstate mixture remains the same.  

Consider a gas to be formed by atoms initially localized at the width of those in a solid, $d$.  This might arise by forming the gas from evaporation or by contact and ``collapse'' of each gas atom during contact with the solid.  We know that such a localized set of parcels, if spreading much slower than the collision time with the particles will tend to the M-B distribution $f(p)$.  The many body wavefunction for such a state can be described by $\Psi\approx \hat S \prod_{i}\psi(x_{i}-R_{i},v_{i}^{MB})$ where each $\psi$ is a localized parcel at $R_{i}$ with nearly monochromatic wavevector $k=mv_{i}\hbar$.  This is certainly a transient state as the packets will delocalize and the packets will spread into a range of directions and speeds as they delocalize.  Nevertheless, the states can be reasonably well represented on the many body basis of eigenstates $\Psi_{k_{1}\ldots k_{N}}(X)$ where the $\{k_{i}\}$ are the one body free wavevectors.  The amplitude near the hard sphere potentials is incorrect but, at low densities and the high energies associated with room temperatures, the error is very small.  The M-B distribution of plane waves corresponding to temperature $T$ gives an eigenstate of net energy per particle $E(T)$ but has no fluctuations.  In contrast, our state will rapidly delocalize and become highly correlated but the persistent feature is that the energy spread per particle, $\Delta E\approx \hbar^{2}/2m d^{2}$, remains the same.  

One may compute the distribution of kinetic and potential energy in a classical gas of CM localized particles using the atom's as approximated by their electron wavefunctions.  This gives a time averaged probability distribution $h_{KE}(E)\approx A$exp$(-(\gamma kT-E)^{2})$ and $h_{PE}\ll h_{KE}$ as $h_{PE}(E)\approx h_{KE}(E) \tau_{r}/\tau_{coll}$ where $\tau_{r}\approx E_{th}/\kappa v_{th}$ is the duration of the recoil which is a function of the atomic shell ``stiffness'' $\kappa$, and the mean thermal velocity $v_{th}$.  These distributions are normalized so that $\int h_{KE}(E)dE=\int h_{PE}(E)dE=1$ and satisfy the energy conservation identity $\int E\cdot h_{KE}(E)+\int E\cdot h_{PE}(E)=E_{th}$. This is in contrast with that of a harmonic solid where potential and kinetic energies are equally shared $h_{KE}=h_{PE}$.  As delocalization of the packets occurs the intensity of the local currents weakens and we can expect a more equal distribution of kinetic and potential energy of the atoms.  

The usual paradox of quantum thermalization gives no equilibration to this state even when we allow for true eigenstates that include the many body potential exactly.  Thermalization must be a universal event.  We have made the argument that starting with M-B packets we attain some universal limit but have not provided a mechanism for more general systems with greater energy spreads to arrive there.  Below we will see that the internal photon field provided such a mechanism and leads to results consistent with the microcanonical ensemble.  This requires oscillations of the electron clouds to generate the electromagnetic fields and the localized packet limit thus the importance of knowing the distribution of kinetic and potential energy distributions (since it is only by variation of the atom's potential energy and its rate that photons are produced).  Interestingly, the electron clouds should reach some thermal level of excitation at finite $T$.  This will alter the binding strength of condensed matter and may provide some insight into the origin of the liquid state which has always been hard to understand in the Born-Oppenheimer approximation.

For our solids described above, if the excitations of the local peaks are large compared to the energy levels of the lower phonons, the peak can move around in a billiard like fashion in this subset of the potential and rapidly give a time averaged filling of this part of the potential well up to the energy surface $E=<E>$ that is traversed on very short time scales.\footnote{This seems superficially to be the case of high phonon occupancy of the lower modes yet the distinction is in the superpositions necessary to produce such states.  As mentioned, the large effective masses of such lower energy modes means the quasi-classical motion of such peaks can persist for long times.}  This localization does not dissipate rapidly like a gas due to its constant coupling to the large mass of the net body and provides a mechanism for equilibration.  Note that this localization property being preserved is regularly observed.  We often think of sound waves as made of phonons yet the acoustic displacements of parts of the solid are often many atomic spacings wide.  The phonon basis is built on small displacements of the peaks from their equilibrium locations not their relative locations to their neighbors.  Therefore, such states are only describable on the phonon basis with broad energy superpositions that can preserve the localization property over displacements far outside the range of validity of the phonon basis.  Interestingly, the dispersion relation for the phonons is the same as for sound waves even at this scale.  

One aspect of thermalization often not discussed in the context of matter is that of the interaction with radiation.  It can be argued that the material coordinates tend to massively dominate the energy distribution so the radiation can be ignored in the thermalization process but electromagnetic waves are essentially noninteracting so never thermalize on their own.  Matter is required to create the thermal radiation field inside matter and to radiate from its surface.   Current fluctuations of a superposition cause compression of the electron orbitals so create photons.  This internal photon field, if universal, should be the Planck distribution.  
Interestingly, this distribution, the original quantum problem, has no dynamical explanation and has never been justified by better than by a ``count the states in a box'' approach.  

Larger energy distributions in a gas or block eigenstates give larger fluctuations which pump more energy into the internal photon field.  Assuming the photons can redistribute the energy nonlocally in the medium we should arrive at a set of wavefunctions in the tower $\Psi_{N},\Psi_{NA},\Psi_{NAA},\ldots$ where the number of $A$'s are the number of photons in the space and the notation is from the reformulation of QED that uses classical fields on a tower of increasing dimensional spaces \cite{Chafin-QED}.  


The fact that photons are so cheap and prolifically produced at low energy has been argued to be the origin of a sparse set of semi-classical wavefunctions in Fock space made of condensing solid clusters that give quantum measurement results in a many-worlds like fashion for long times \cite{Chafin-QED}.  We suggest that these fluctuations generate the Planck distribution and that the fluctuations in it are due to local source inhomogeneities (thus a non-universal part that dissipates with distance from the interacting matter) and from the field variations in the relative phases of the different photon number spaces (a universal component).  



\section{Planck Distribution and Fluctuations}

Consider a medium that is optically opaque on a distance of many atomic spacings.  We are seeking an equilibrium state between the oscillating material flux of $N$ particles and the photons in a Fock space tower of functions we denote by $\Psi_{N},\Psi_{NA},\Psi_{NAA}\ldots=\Psi_{N},\Psi_{N,1},\Psi_{N,2}\ldots$.  These are functions that exist on large dimensional spaces.  Explicitly, $\Psi_{N}(X)=\Psi_{N}(x_{1}\ldots x_{N})$, $\Psi_{Nm1}(X)=\Psi_{N}(x_{1}\ldots x_{N},x_{1}^{(A)})$, etc.\ where spin and vector indices have been suppressed.  An atomic spin label should be included with every coordinate variable and an vector index should be included with every $x^{(A)}$ coordinate.  
 The oscillations driven by inhomogeneity and the spread of eigenstate energy drives photon creation up the tower.  When there is enough electromagnetic energy in a mode to balance the flux this stops.  In any given pair collision the energy and relative orientation varies but will tend to follow some equilibrated distribution.  This generates a corresponding set of charge oscillations and induced photon frequencies as a function of the duration of contact and size of the deformation during the collision.  However, this is not the same as black body distribution which we know is universal as a function of $T$ and not dependent on the particular mass and shape of the atoms that generate it.  

To understand the origin of the thermal photon distribution we should start with the photon distribution created by a pair collision in a photon free space.  There is a distribution of photons $f(\omega)$ created about some peaked frequency $\omega_{0}$.  In the case of a system that starts with no photons, this drives amplitude and energy from $\Psi_{N}$ to $\Psi_{N,A}$.  As $\Psi_{N,A}$ gains amplitude the radiation flows back into $\Psi_{N}$.  The impulses of this radiation are negligible for free atoms since photons carry very little momentum relative to their energy but during collisions it can impart energy and cause the two atoms to fly apart with increasing velocity.\footnote{This is a point easily missed for those dedicated to diagrammatic methods where the momentum conservation laws at vertices conserve pseudomomenta and are not related to conservation of true momentum.}  At equilibrium, a variation in the photon energy eigenstate distribution creates spatial currents and spatial variations in field strength.  This is passed on to the matter which then dissipates such variations down to a scale comparable to the free path.  As a result we expect the photon sector of each $\Psi_{N,m}$, when we fix the material coordinates, to be a near eigenstate, a random member of the eigenstate sector corresponding to some photon energy $E_{\gamma}(m)$.  Since thermal photon wavelength are often much longer than interparticle separations, we can expect an energy spread in the photon spectrum to be less than one typical wavelength.  
\begin{gather}\label{tower}
\vdots\\\nonumber
\Psi_{N,n}\\\nonumber
\vdots\\\nonumber
\Psi_{N,4}\\\nonumber
\Psi_{N,3}\\\nonumber
\Psi_{N,2}\\\nonumber
\Psi_{N,1}\\\nonumber
\Psi_{N}\nonumber
\end{gather}

The creation of photons is dominated by single electron processes of accelerating electron current about its associated atomic core.  As two adjacent atoms are pressed together with phonon fluctuations or atoms in gas collisions, photons can be created.  A similar situation occurs for photon absorption.  In our photon tower of states the zero photon state $\Psi_{N,0}$ radiates into the state $\Psi_{N,1}$ and similarly the electromagnetic field oscillations in the latter can drive energy and amplitude into the former.  

We assume a weak coupling between photons and atoms to make the problem tractable so that assuming product functions of matter and photons for eigenstates is reasonable.  Combined with the above result we can assume our eigenstates are near to the form $\Psi_{N}^{(m)}\mathcal{A}_{m}$ where $\mathcal{A}_{m}$ is a stationary state in the space spanned by $A^{i_{1}}\otimes A^{i_{2}}\otimes\ldots\otimes A^{i_{m}}$ of complex 3-vectors fields for photons.\footnote{Coulomb gauge is assumed for every coordinate label so that the $\Psi_{N,1}^{\mu=0}$, $\Psi_{N,2}^{\nu,\mu=0}$, etc.\ components are fixed by constraint.}  Normalization of objects in this space is given by the complex KG norm $(4\mu_{0})^{-1}\Im (A^{i}\dot A^{i,*}-\dot A^{i,*} A^{i})$ for evolution along one of the photon time directions, $t^{A}_{i}$, \cite{Chafin-QED} so that photons of the form $A=\alpha e^{ik\omega}$ have norm $\alpha^{2}\omega/2\mu_{0}$.  Norm conservation is summarized by the result: $\sum_{i} \hat{\mathcal{N}}(\Psi_{N,i})=1$ where $\hat{\mathcal{N}}$ is the norm operator for the associated matter-photon field.  For example,
\begin{align}
\hat{\mathcal{N}}(\Psi_{N,n})=\frac{1}{4\mu_{0}}\int dx^{i_{1}}_{s}\ldots dx^{i_{N}}_{s} \int dx^{i_{1}}_{A}\ldots dx^{i_{n}}_{A}\sum_{k=1}^{n} \left( \Psi^{i_{1}\ldots i_{n}}\partial_{t_{A}^{i_{k}}}\Psi^{*}_{{i_{1}\ldots i_{n}}} -\partial_{t_{A}^{i_{k}}}\Psi^{{i_{1}\ldots i_{n}}}~\Psi^{*}_{{i_{1}\ldots i_{n}}}\right)
\end{align}
This is the norm of the function corresponding to $N$ atoms (labeled with $s$) and $n$ photons where the $i$ superscripts range over 1, 2, 3 for each rectangular coordinate direction and repeated indices are summed.  The many time labels allow derivatives to be taken with respect to each of them.  This is a conserved quantity as long as we remain on the equal times diagonal $t^{i_{1}}_{s}=t^{i_{2}}_{s}=\ldots=t^{i_{1}}_{A}=t^{i_{2}}_{A}=\ldots$.  Such an invariant opens up a connection between the formalism of quantum field theory and a well posed initial value formalism.  By looking at acceleration of currents in one space and the change in energy and norm in the next higher number photon space one can extend radiation reaction methods to include both 4-momentum and norm conservation.  By ``current'' here we are not looking at, for example, $j(\Psi_{e})$ to get the evolution in the $A^{i}$ time direction of $\Psi(eA)$.  Rather, the ``current'' is implicitly defined by the relative amplitudes of $\Psi_{e}$ and $\Psi_{eA}$ together.  For example, one can locally write $\Psi_{eA}(x,y)\approx \Psi_{e}'(x)A^{i}(y)$.  The current that defines the evolution in the $A$ direction is that of $j=i\bar\Psi\gamma^{i}\Psi'$.  It should be reemphasized that if there is no current change there is no radiation field and no norm change in the $(N-1)$th and $N$th photon number states.\footnote{One can have oscillating photon states with $A^{i}$ being completely real that give no norm contribution in our tower however, such states are decoupled from the state of systems given by our initial data of macroscopically colliding zero temperature blocks.  To generate the vector potential contributions one must transfer both norm and energy up the tower one must satisfy constraints on the current and norm fluxes at the interacting ``diagonals.''  For plane wave production this ratio is $j_{E}/j_{norm}=\hbar\omega$.}  

The norm is conserved over the tower of states but this is also a crucial concept in defining the conservation of energy over this tower.  We can define the one particle energies of the material and photon labels by their free field Lagrangians.  These definitions also hold if we evolve our many particle functions along a \textit{single} particle time coordinate level.  Of course, this is not what is physically relevant.  We need a definition of energy that is conserved for the whole tower when evolved along the \textit{equal times}.  Let us define the single particle energy functions as $\hat{E}_{s}$ for the material coordinates and $\hat E_{A}$ for the photon ones.  The conserved equal times energy is then 
\begin{align}
E_{N,k}=\bar\Psi_{N,k}\left(\sum_{i=1}^{N}\hat{E}_{s_{i}}\hat{\mathcal{N}}_{1\ldots \hat i \ldots N}\hat{ \mathcal{N}}^{A}_{1\ldots k}+\sum_{j=1}^{k}\hat{E}_{A_{j}}\hat {\mathcal{N}}_{1\ldots N}\hat {\mathcal{N}}^{A}_{1 \ldots \hat j \ldots k}\right)\Psi_{N,k}
\end{align}
where $\bar\Psi_{N,k}$ is defined as $\gamma^{0}_{ab}\gamma^{0}_{a'b'}\ldots \bar\Psi_{N,k;bb'\ldots}^{*}$ so that all the spinor indices are contracted with a $\gamma^{0}$
The net energy of the system is then $E=\sum_{k}E_{N,k}$.  The net norm is similarly $\sum_{k}\hat{\mathcal{N}}(\Psi_{N,k})=1$.

Since the photon fraction of energy is generally very small compared to the material contributions we can assume that the persistent (time varying) density fluctuations, from having a finite energy spread in our original photon-free eigenstate distribution, persist after equilibration in the photon Fock space tower.  Thermalization suggests that we approach some universal limit in this case.  This forces us to confront the paradox of the persistence of eigenstate distributions in quantum mechanics.  It is possible that the details of the radiation reaction has introduced some nonunitary evolution in the dynamics but this author is highly skeptical of this being the driving factor.  Our earlier discussion of solids and the persistent localization of atomic centers as we ``heat'' the material by macroscopic actions suggests that a similar mechanism may arise here.

In the case of a solid we now have reason to believe that the atomic cores have well defined location and that this persists for all temperatures.  We have not defined temperature yet but have suggested that it corresponds to a universal limit of short time equilibrations of motion for the class of excitations that begin as macroscopic changes.  For a block of atomic matter with negligible photon effects the long time equilibration suggests that the limit is well approximated for most of its history by the microcanonical ensemble i.e. narrow energy width random mixtures of eigenstates.  Of course, the presence of fluctuations and localization means that the true state is not such a narrow distribution and the energy spread can be very large.    

Now allowing for photon creation the norm and energy will distribute itself among the tower in eqn.\ \ref{tower}.  An aspect of the fact that the norm of a photon is frequency dependent is that low energy photons are prolific.  As such we expect even very low temperatures to produce many more photons, i.e. to fill our tower up to levels far higher, than the number of atoms $N$.  If we have spatial variations in the material density then this enhances the photon production there and, since photons can travel many interparticle separations, this has an equilibrating effect in that the spatial density of energy and density tends to uniformity.  However, the localizations of the atomic cores should nevertheless remain well defined throughout this tower since the flux of norm that fills the towers carries with it these locations and they remain localized for the same reasons as before.  

For this to persist we should have a narrow spread of energies even among these excitations.  As such we expect the energy within each occupied $\Psi_{N,i}$ to be the same.  (More precisely, we should expect the energy/norm of each such state to be the same.)  Since the $\Psi_{N,k}\approx \Psi_{N}^{k}\mathcal{A}_{k}$ function exists as a well defined peak and velocity in the 3N dimensional harmonic well and this is argued to be the same for every $k$.  This is just equivalent to a classical phase space harmonic oscillator $\prod(p_{i},q_{i})\in\mathbb{R}^{6N}$ (suppressing spin labels).  The photon field for each $k$, $\mathcal{A}_{k}$, lives in a space spanned by the products of $e^{ikx}\hat\epsilon_{j}$ vectors of unit photon norm.  $\Psi_{N,k}$ then exists as a point in $\mathbb{R}^{6N}\otimes \prod\mathbb{R}^{3k}$.  The general tower state is a point in $\mathbb{R}^{6N}\otimes \sum_{k} \prod\mathbb{R}^{3k}$ of fixed energy and norm.  For a general incommensurate orbit, we expect it to spend most of its time on the constant norm and energy (and other constant conserved integral quantities) hypersurfaces where it is dominated by ``typical'' values.  This justifies the extremely peaked distributions that favor microcanonical occupancy statistics for the photons.  
%

This gives a sharply peaked distribution for each about the same value of $T^{-1}\doteq \partial \ln g/\partial E$ where $g(E)$ is the density of energy eigenstates near to $E$.\footnote{This was essential for self consistency.  If there was a large spread in the energy of the massive field contributions the localization property of the atomic peaks would fail when different photon number spaces are compared.}  Note that for any given system we expect it to be a roughly equal distribution among the various photon number states $k$ and to be in one particular state $\Psi_{N,k}$ for each of them, certainly \textit{not} an eigenstate but with some time averaged properties of one.  This is how the notion of temperature arises based on energy eigenstate density becomes relevant for the dynamics of the general wavefunction that arises from our ``classical'' initial data.  The fluctuations are essential in producing the kind of universality in long time limits we associate with temperature and equilibration and yet are irrelevant in giving the particular distribution which is, ironically, determined by the eigenstates best associated with the time averaged states.  This gives both a dynamical explanation of the Planck distribution and a physical origin for the relevance of the calculations that arise from the microcanonical ensemble without its evident contradictions with quantum dynamics.  
 \begin{figure}
 \begin{centering}
 \includegraphics[width=2in]{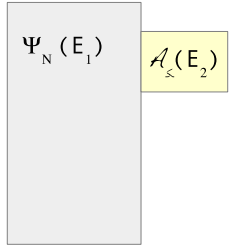}%
 \caption{\label{coupling}Each fixed photon number function in the tower can be represented as a material part and a photon field part in a manner reminiscent of couplings of quantum thermodynamic systems in an ensemble.  The apparent dominance of eigenstate contributions to the states given general physical systems is explained in the text.}%
  \end{centering}
 \end{figure}

\section{Conclusions}
The thermalization of a solid has been shown to be a well defined universal limit for a particular subset of wavefunctions that correspond to recognizable macroscopic matter that has arisen from a very cold initial state.  This ``limit'' must be interpreted to be over a long but not infinite time scale so that inevitable delocalization properties that are essential both for its ``classicality'' and internal localizations that create the driving time dependent fluctuations dominate.  

The quantum statistical ensembles have always been a deep mystery.  Inspired by classical mechanics they have provided the framework for many successful calculations with no inspiration for why they work.  Quantum theory has been rife with such ``formal procedures'' out of necessity but has left us with many problems with no intuition to move forwards.  It will be both ironic and exciting if these ensembles give far less generally valid result in extreme but apparently ``equilibrated'' matter.  The way forwards on such systems seem like it must arise from quantum dynamics itself so that continuing to treat thermal properties by a separate set of rules from the fundamental dynamics itself must be a dead end.  

There is clearly much more work to do if such perspective is to be fully justified, and even if it is, many new doors to problems will open that will require their own internal consistency checks.  The case of gases is only cursorly considered here.  Gas atoms have a delocalization that is inevitable on short scales and, even in their interactions with condensed matter, drive them back to quasi-localized behavior, the bulk of the gas has a very long journey to make to such solid interfaces so that this seems unlikely to dominate the dynamics.  The interaction of gases and the internal thermal photon field seems crucial for the understanding of topics such as nucleation theory and other fields still poor in finding agreement of theory and experiment.  In the case of phase transitions, fluctuations play a very large role and the time dependence of them matters.  The usual quantum statistical formalism does not give explicit answers or even a typical state to evolve to probe this.  In the case of liquids, localization should persist but the binding between atoms in the Born-Oppenheimer approximation is very rigid.  An approach that allows for thermal excitation of the electron orbitals offers a way to reduce the angular barrier energy of three-body clusters and allow a more rapid diffusive flow of neighbors relative to each other.

Ultracold gas experiments have been analyzed through hydrodynamic and thermodynamic constructions with field theoretic and Monte Carlo methods to derive the relevant physical parameters \cite{Dolfovo}.  Great interest has been paid to the thermodynamic phase portrait of these fermions especially at ``unitarity'' in the hopes that it will give scalable results for higher energy systems.  It is this author's opinion that such gases are too isolated and delocalized to have any well defined notion of temperature and that the kinds of delocalization that exists makes the use of hydrodynamic parameters to describe velocity and density in a 3D fashion meaningless.  It would be very interesting to see if evaporative cooling could be conducted so that we can vary the net energy of the cloud and the variation in the energy eigenstates that make it up independently.  This could allow us to vary the size of the cloud and the size and rate of its fluctuations independently, thus foiling any thermodynamic approaches for these systems.  Thermodynamics and hydrodynamics presume some universal limit that is independent of the origins of the system beyond such macroscopic and parcel averaged parameters.  If these systems hold such history dependence then our next task should be to classify and quantify its manifestation from a broader perspective.  Properties such as viscosity and entropy, eagerly sought in the quest for a universal quantum bound on hydrodynamics \cite{Son}, may turn out to be meaningless and a great deal of work on fermionic phases and such limits will need to be reconsidered.  



\begin{thebibliography}{9}

\bibitem{Bartsch} C. Bartsch and J. Gemmer, Dynamical Typicality of Quantum Expectation Values, Phys. Rev. Lett, {\bf 102}, 11, (2009). 

\bibitem{Boltzmann}
L. Boltzmann,
\newblock {\em Lectures on Gas Theory}.
\newblock University of California Press, english edition edition, 1964.

\bibitem{Chafin-I}
C. Chafin,
The Quantum State of Classical Matter I: Solids and Measurements, quant-ph/arXiv:1308.2305.

\bibitem{Chafin-II}
C. Chafin,
The Quantum State of Classical Matter II: Thermodynamic Equilibrium and Hydrodynamics (2014), quant-ph/arXiv:1309.1111.

\bibitem{Chafin-QED} C. Chafin, Beyond Quantum Fields: An Operator-Free Covering Theory for QED, (2014), under review.  

\bibitem{Chafin-Qmeasurement}
C. Chafin, 
Quantum Measurement and Classical States, (2014), quant-phys/arxiv.org/abs/1410.8238.

\bibitem{Cohen} G. Gallavotti, E. G. D. Cohen, Dynamical ensembles in nonequilibrium statistical mechanics, Physical Review Letters, 74, 2694--2697, 1995; and Dynamical ensembles in stationary states, Journal of Statistical Physics, 80, 931--970, 1995,

\bibitem{Dolfovo} F. Dalfovo, S. Giorgini, L. P. Pitaevskii, S. Stringari,
Theory of Bose-Einstein condensation in trapped gases,
Rev. Mod. Phys., {\bf 71}, 3, 463, 1999.

\bibitem{Einstein} A. Einstein, (1917), Phys. Zeit. 18, 121 (1917). Reprinted in the Collected Papers of Albert
Einstein, Engl. Transl. by Anna Beck, Princeton, vol 6.

\bibitem{EinsteinH} A. Einstein and L. Hopf (1910), Ann. Physik 33, p 1105. Reprinted in the Collected papers of
Albert  Einstein. Engl. Transl. by Anna Beck, Princeton, vol 3.

\bibitem{Jarz}
C. Jarzynski, "Nonequilibrium equality for free energy differences"
\newblock Phys. Rev. Lett. {\bf 78}, 2690 (1997).

\bibitem{Jeans}
J. H. Jeans.{\em The Dynamical Theory of Gases}, The University Press, (1904).

\bibitem{Popescu} S. Popescu, A. J. Short, and A. Winter. Entanglement and the foundations of statistical mechanics - 2006. Nature Phys., {\bf 2}, 754.

\bibitem{Reif} F. Reif,  Fundamentals of statistical and thermal
 physics, Singapore: McGraw-Hill, 1965.
 
\bibitem{Son}
  D.~T.~Son and A.~O.~Starinets, Viscosity, Black Holes, and Quantum Field Theory,
  Ann.\ Rev.\ Nucl.\ Part.\ Sci.\,  {\bf 57}, 95, (2007),
  [arXiv:0704.0240 [hep-th]].

\bibitem{Srednicki} M. Srednicki, Chaos and Quantum Thermalization. Physical Review E {\bf 50} (2): 888,  (1994).

\bibitem{Srednicki1} M. Srednicki, The approach to thermal equilibrium in quantized chaotic systems. Journal of Physics A: Mathematical and General {\bf 32} (7): 1163,  (1998).

\bibitem{Villani} L. Desvillettes and C. Villani, On the Trend to Global Equilibrium for Spatially Inhomogeneous Kinetic Systems: The Boltzmann Equation, Comm. Pure Appl. Math, {\bf 54},1, (2003).

\end{thebibliography}
\end{document}